\documentclass[reprint,amsmath,amssymb,aip,jmp]{revtex4-1}

\usepackage{bmpsize}
\usepackage{graphicx,color}								
\usepackage{amssymb}
\usepackage{natbib}
\usepackage{array}
\usepackage{subfigure}
\usepackage[usenames,dvipsnames,svgnames,table]{xcolor}

\newcommand{\paddyspeaks}[1]{{\color{magenta} #1}}
\newcommand{\BE}{\beta\varepsilon}
\newcommand{\tp}{\tau_{\rm perc}}
\begin{document}

\title{Local structure of percolating gels at {very} low volume fractions}

\author{Samuel Griffiths}
\affiliation{School of Chemistry, University of Bristol, Cantock's Close, Bristol, BS8 1TS, UK}
\author{Francesco Turci}
\email[Corresponding author: ]{f.turci@bristol.ac.uk}
\affiliation{H.H. Wills Physics Laboratory, Tyndall Avenue, Bristol, BS8 1TL, UK}
\author{C. Patrick Royall}
\affiliation{School of Chemistry, University of Bristol, Cantock's Close, Bristol, BS8 1TS, UK}
\affiliation{H.H. Wills Physics Laboratory, Tyndall Avenue, Bristol, BS8 1TL, UK}
\affiliation{Centre for Nanoscience and Quantum Information, Tyndall Avenue, Bristol, BS8 1FD, UK}

\begin{abstract}
The formation of colloidal gels is strongly dependent on the volume fraction of the system and the strength of the interactions between the colloids. Here we explore very dilute solutions by the means of numerical simulations, and show that, in the absence of hydrodynamic interactions and for sufficiently strong interactions, percolating colloidal gels can be realised at very low values of the volume fraction. Characterising the structure of the network of the arrested material we find that, when reducing the volume fraction, the gels are dominated by low-energy local structures, analogous to the isolated clusters of the interaction potential. Changing the strength of the interaction allows us to tune the compactness of the gel as characterised by the fractal dimension, with low interaction strength favouring more chain-like structures. \end{abstract}

\maketitle
\section{Introduction}

{When subject to a moderate quenching, a large variety of systems can form macroscopic networks of arrested materials, also called \textit{gels} \cite{zaccarelli2007,poon2002,coniglio2004,ramos2005}. Systems as different as proteins \cite{tanaka2005}, clays \cite{jabbarifarouji2007}, foods \cite{tanaka2013fara}, hydrogels \cite{helgeson2012} and tissues \cite{drury2003,rose2014} can undergo gelation, with innumerable applications, as well as more exotic kind of systems such as phase-separating oxides \cite{bouttes2014} and metallic glassformers \cite{baumer2013}.  }


{In order to predict the mechanical properties of gels, it is important to know both their local\cite{zaccone2011,zhang2012,hsiao2012} and global \cite{valadezperez2013,kroy2004,varrato2012} structure, but a deep understanding of both remains today a challenge.}
 For example, in the very dilute limit, the study of gel formation via molecular dynamics is challenged by the very long times required to form aggregates, with equilibration times that easily exceed $10^8$ integration steps \cite{sciortino2004}. 
In the model colloidal gels we will consider, demixing of the particles into a (colloidal) ``gas'' and ``liquid'' occurs. Spinodal demixing leads to a network of particles 
\cite{verhaegh1997,tanaka1999colloid,manley2005spinodal,lu2008,zaccarelli2008} which undergoes dynamical arrest \cite{testard2011,zia2014}. The final structure can persist for years  \cite{ruzicka2010prl}, if the self-generated or gravitational stress is weaker than the yield stress \cite{tanaka2013fara,cates2004theory}.  Demixing is driven by effective attractions between the colloidal particles induced by the addition of non-absorbing polymer. Thus, although the {original} system is a mixture of three important components --- colloids, polymers and solvent --- {we can build an effective one-component model} of colloids which experience a {pair, spherically symmetric} attractive interaction whose strength {corresponds to different} polymer concentrations \cite{dijkstra2000,likos2001}. 

\begin{figure}[t]
	\centering
	\includegraphics{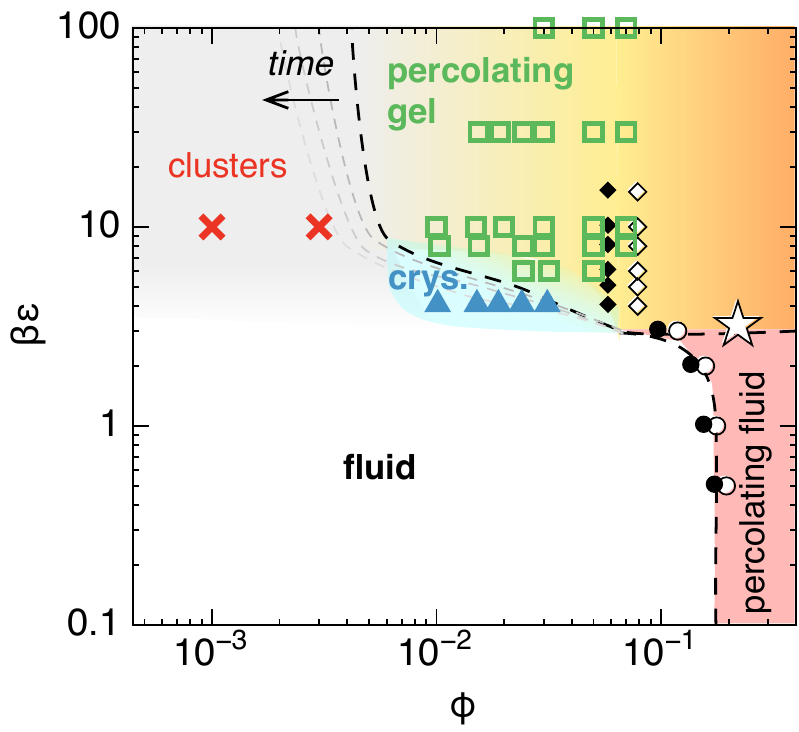}
	\caption{Simulation state points on a schematic phase diagram for gels modelled by Morse interactions and Brownian dynamics {(notice the double logarithmic scales)}: black and white symbols are from \cite{royall2015}, {and identify the previous estimates for the percolation transition}. Circles identify the boundaries between fluid and percolating fluid, while black and white lozenges correspond to cluster and gel phases, when the observation time is chosen to match the experimental conditions\cite{royall2015}. {The star corresponds to the approximate position of the critical point \cite{loverso2006}.}
	 With the present work, {we explore the states indicated by squares, crosses and triangles}: we follow the system till percolation occurs, and find gels down to much smaller volume fractions (green squares) {than previously observed}. Non percolating clusters (red crosses) are found when the percolation time $\tp$ exceeds the accessible computer simulation time: {the percolation line (dashed line) is time-dependent and when time increases, it moves to lower volume fractions (gray-shaded dashed lines)}. {In a limited range of interaction strengths,} we also observe the phase separation into gas and crystalline {``droplets''} (blue triangles).
	}
	\label{diagram}
\end{figure}

Nonetheless, the spinodal decomposition scenario is not the only possible mechanism, and due to some discrepancies in the literature \cite{miller2003,lu2008} alternative pathways to gelation have been proposed \cite{eberle2011} based on percolation. 

In the present work we investigate the low-volume fraction limit of gelation  (neglecting hydrodynamic interactions and their long-range, multi-body effects \cite{furukawa2010,royall2015}) and its relation with percolation for a system of Brownian particles with tuneable short-range attractions. In particular, we demonstrate that the structural features present at moderately low {volume fraction} $\phi\approx 10^{-1}$ survive at much lower {volume fractions}, $\phi\approx10^{-3}$, and allow for the formation of thin percolating structures. When the interaction strength is strong enough (\textit{i.e.} when the effective temperature is low enough), we show that aggregation proceeds systematically and that a power law behaviour relates the increase in the time to form a percolating gel and the inverse of the volume fraction. {Thus gelation from a state of clusters --which may be extended in space-- may be viewed as a time-dependent percolation transition.}	
	
The article is organised as follows: in section \ref{sec:model} we present the numerical model and the protocol followed in our numerical simulations; in section \ref{sec:metric} we illustrate the observables used to characterise the structure of the simulated gels; in section IV we report the main results concerning the formation of nonequilibrium gels at low volume fractions and their structural features, followed by the analysis of non-percolating clusters at {very} low densities; we conclude with a critical assessment of our results.

\begin{figure*}[t]
\centering
\includegraphics{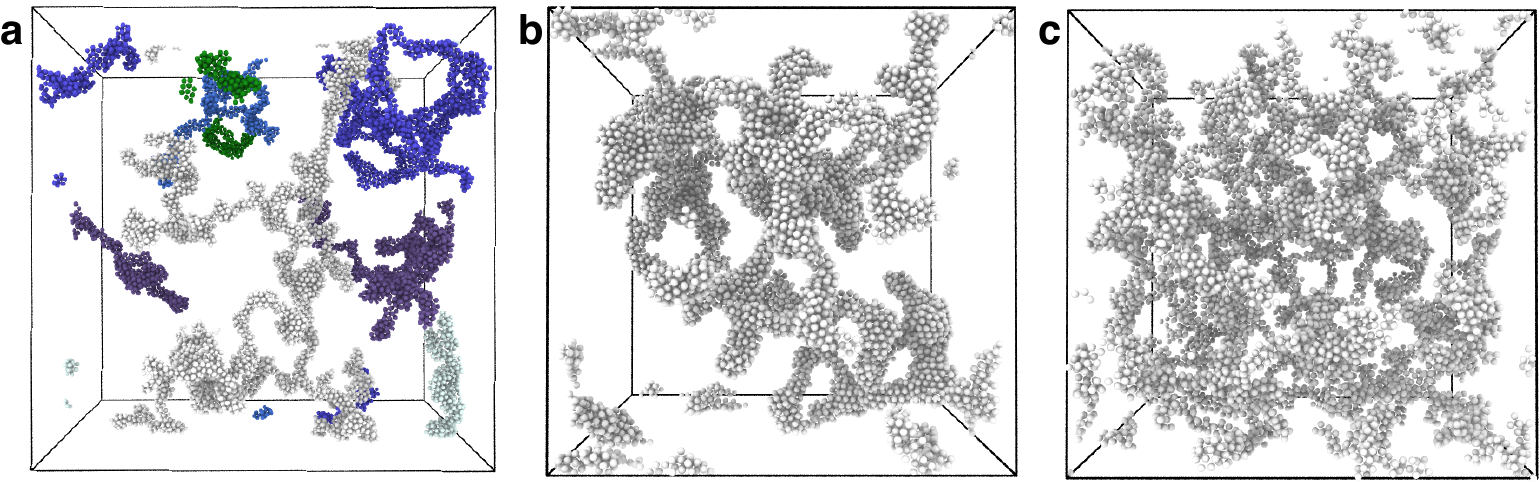}
\caption{Aggregates (coloured) and percolating clusters (white particles) for different volume fractions and interaction strengths:  (a) $\phi=0.015$ and $\beta\varepsilon=10$,  (b)$\phi=0.05$ and $\beta\varepsilon=6$, (c) $\phi=0.05$ and $\beta\varepsilon=100$ .}
\label{fig:percs}	
\end{figure*}

\section{Model}
\label{sec:model}
We perform molecular dynamics simulations \cite{plimpton1995} of a model gel based on simple interaction potentials. We consider a polydisperse additive mixture of particles of different diameters. Particles $i$ and $j$ interact via a truncated and shifted Morse pair potential $u(r)$
\begin{equation}
\beta u(r_{ij}) = \beta\varepsilon \, \exp[\rho_{0}(\sigma_{ij} - r_{ij})](\exp[\rho_{0}(\sigma_{ij} -r_{ij})]-2),
\label{eq:morse}
\end{equation}
where $\beta=1/k_BT$ is the inverse temperature with Boltzmann constant $k_B$, $\rho_0=33$ is the range parameter and $\sigma_{ij}=(\sigma_i+\sigma_j)/2$. The interaction potential is truncated at distance $r_{\rm cut}=1.4\sigma_{ij}$. Particle sizes are drawn from a Gaussian distribution of mean $\sigma$ and width $\Delta$, with polydispersity $\Delta/\sigma=4\%$.  This effectively reproduces the physics of colloid-polymer mixtures, leading eventually to gelation due to the very short {interaction} range and very strong attraction, whose amplitude is determined by the interaction strength $\beta \varepsilon$ \cite{royall2008,royall2008aip,taffs2010,royall2015}.

We consider systems of $N=10\,000$ and $100\,000$ particles performing over-damped Langevin dynamics. We explore a wide range of interaction strengths $\beta\varepsilon$ and volume volume fractions $\phi=\frac{1}{6L^3}\sum_i^{N}\pi\sigma_i^3$, where $L$ is the linear size of the cubic simulation box, with periodic boundary conditions. In particular we focus on very small volume fractions, down to $\phi=0.001$, and very strong interactions, up to $\beta\varepsilon=100$. While these may exceed the interaction strengths typically associated with colloid-polymer mixtures, van der Waals interactions between colloidal particles are of this order and greater \cite{russel}.

Every simulation run starts from an initial random distribution of the particle centres. Velocities are then randomly assigned from a Maxwell-Boltzmann distribution at inverse temperature $\beta$ and the particles undergo an over-damped Langevin dynamics with Brownian time $\tau_B=(\sigma/2)^2/6D$ where $D$ is the self-diffusion constant for a particle of diameter $\sigma$ and it is related to the friction coefficient $\gamma$ by Stokes's law $D=1/\beta\gamma$.
. We integrate the equations of motion using the velocity-Verlet algorithm with time-step $dt=0.001\sqrt{m\sigma/\varepsilon}$  and $\gamma=10 \sqrt{m\varepsilon/\sigma}$, evolving the system for a maximum of $2\cdot10^9$ integration steps. Average values and standard errors are evaluated from 6 distinct trajectories for every state point. Throughout we employ the LAMMPS molecular dynamics package \cite{plimpton1995}.

{The state points that we consider} are represented on the schematic diagram in Fig.~\ref{diagram}: most of the simulations have been run in the percolating gel phase (green squares), with volume fractions in the $[0.01,0.7]$ interval, as well as the formation of crystalline aggregates (blue triangles); we also sample {very} low volume fractions (red crosses) where a {non-crystalline} cluster phase is observed. We focus in the following on the relevant structural features that distinguish these different phases.
{The gelation region (where the system has a sufficient interaction strength $\BE$ to undergo de-mixing, meaning that $\BE>\beta_c\varepsilon$ at criticality) is determined through the extended law of corresponding states where the reduced second viral coefficient $B_2^* \approx -1.5$ requires that $\BE \lesssim -2.96$ \cite{noro2000}. }

\section{Metrics}
\label{sec:metric}
In this work, structural measurements on gels and aggregates are performed focusing on two-point correlations (static structure factor)
and higher order correlations as detected by the Topological Cluster Classification (TCC) \cite{malins2013tcc}. 

Emerging characteristic length scales can be identified\cite{foffi2005} computing the structure factor $S(q)$ directly from the pair correlation function $g(r)$ as 
\begin{equation}
	S(q)=1+4\pi\frac{N}{qV}\int_0^{L/2} r \sin(qr)[g(r)-1] dr.
	\label{eq:sq}
\end{equation}

The TCC, instead, provides a library of structures composed of $m$ particles that are ground states (energy minima) for the chosen interaction potential in the case of mono-disperse mixtures\cite{miller1999,doye1997}.

Using the Voronoi network of direct neighbours and selecting particles within the cutoff distance $r_{\rm cut}$ we are able to reconstruct the entire neighbourhood of every single particle and identify whether it is compatible with one or more of the candidate structures of the TCC. In particular, we focus on structures of $m=5,8,9,10$ particles relevant to the Morse interaction and we additionally check for local crystalline order, such as face centred cubic (fcc) and hexagonal close packed (hcp) order. It is important to notice that some of these local minima have an immediate geometric meaning: the $m=5$ triangular bipyramid corresponds to tetrahedral order, the $m=10$ {defective icosahedron} to five-fold symmetric order. As a result of the TCC analysis, we obtain the concentration of structures $n_m=N_m/N$ for every candidate structure, where $N_m$ corresponds to the number of particles belonging to a collection of structures of type $m$. {Since a given particle can be in principle associated to multiple kinds of local order, we choose to establish a hierarchy based on the size of the candidate structure and label the particles consequently: the priority of label assignment follows the order of the list $\{\rm fcc, hcp, 10,9,8,5\}$.}%

\begin{figure}
	\centering
	\includegraphics{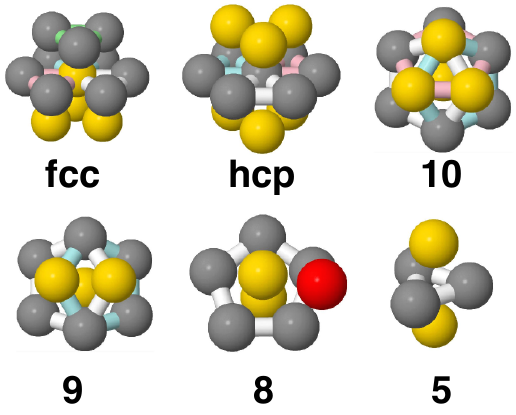}
	\caption{The local energy minima for $m$ identical particles interacting with the Morse potential considered by the TCC algorithm in order to detect local order. Every structure is composed of rings of particles (highlighted in different colors). Structures with $m=8,9,10$ are based on five-membered rings. The $m=5$ structure corresponds to tetrahedral order.  }
\end{figure}

The formation of gels entrains the assembly of extended aggregates of particles that percolate through the simulation box, see Figure \ref{fig:percs}. We call these aggregates \textit{clusters} and we detect them with an agglomerative algorithm \cite{stukowski2009visualization}: particles at a distance lesser than $r_{\rm cut}$ are connected and belong to the same cluster; the system is partitioned into distinct clusters whose maximal extension $\ell_x,\ell_y,\ell_z$ in the $x,y$ or $z$ dimension can be compared with the size of the box. Whenever $\ell_i>L-2\sigma$ we identify the cluster as a \textit{percolating cluster}, and the system is considered to be a gel.

An additional direct estimate of the size of the aggregates is provided by the radius of gyration $R_g$, defined, for an aggregate of $P$ particles, as
\begin{equation}
	R_g=\left\langle \frac{1}{P}\sum_{i=1}^P (\vec{r}_i-\vec{r}_{cm})^2\right\rangle,
\end{equation}where $\vec{r}_{cm}$ is the position of the centre of mass of the aggregate and $\langle\cdot\rangle$ indicates the average over the ensemble of aggregates. The growth of the radius of gyration in time allows to describe the aggregation process.

Finally, we also quantify conformational changes by the means of the fractal dimension of the gel configurations. In particular we estimate the Hausdorff dimension \cite{mandelbrot1983fractal} (naturally smaller than the Euclidean dimension) following the box counting algorithm \cite{gagnepain1986}, which subdivides the system in cells of variable linear size $s$ and evaluates the number of cells $N_c$ filled by the gel's branched structure as a function of $s$. The box-counting fractal dimension is defined as

\begin{equation}
	d_f=\lim_{s\rightarrow 0} \frac{\log N(s)}{\log 1/s},
\end{equation}
 and it provides (in the limit of large systems) an estimate of the fractal dimension $d_f$ of the gel's structure.

\section{Results} 

{We divide our analysis into four parts: we first describe the dynamics of gel formation and its consequences on the large wavelength structural features (Sec.~\ref{subsec:timev}); from this we estimate the time necessary to form percolating networks for lower and lower volume fractions and observe no sign of an intrinsic limit volume fraction for sufficiently large interaction strengths (Sec.~\ref{subsec:limits}); we then focus on the structural features of the percolating networks at low ($\approx 0.01$, Sec.~\ref{subsec:low}) and extremely low ($\approx 0.001$, Sec.~\ref{subsec:xlow}) volume fractions, making use of the higher order correlation functions provided by the Topological Cluster Classification.}

\subsection{Time evolution}
\label{subsec:timev}
Starting from a uniformly distributed initial random configuration, particles diffuse and eventually interact when approaching each other at distances below $r_{\rm cut}$. Due to the deep {attractive} wells of the chosen potential, the system rapidly enters a regime of very slow relaxation, where the particles gradually form clusters whose extension depends on the thermodynamic conditions and on the observation time. The slow relaxation process involves a continuous drift of the potential energy per particle towards lower and lower values, due to the continuous (and progressively slower) reorganisation of the particles within the clusters and the growth of more extended clusters. At a time $\tp$ -- characteristic of the chosen volume fraction $\phi$ and interaction strength $\BE$ -- a large percolating cluster is formed, and we regard the system as a \textit{nonequilibrium percolating gel} \cite{zaccarelli2007}.

\begin{figure}[t]
	\centering
	\includegraphics{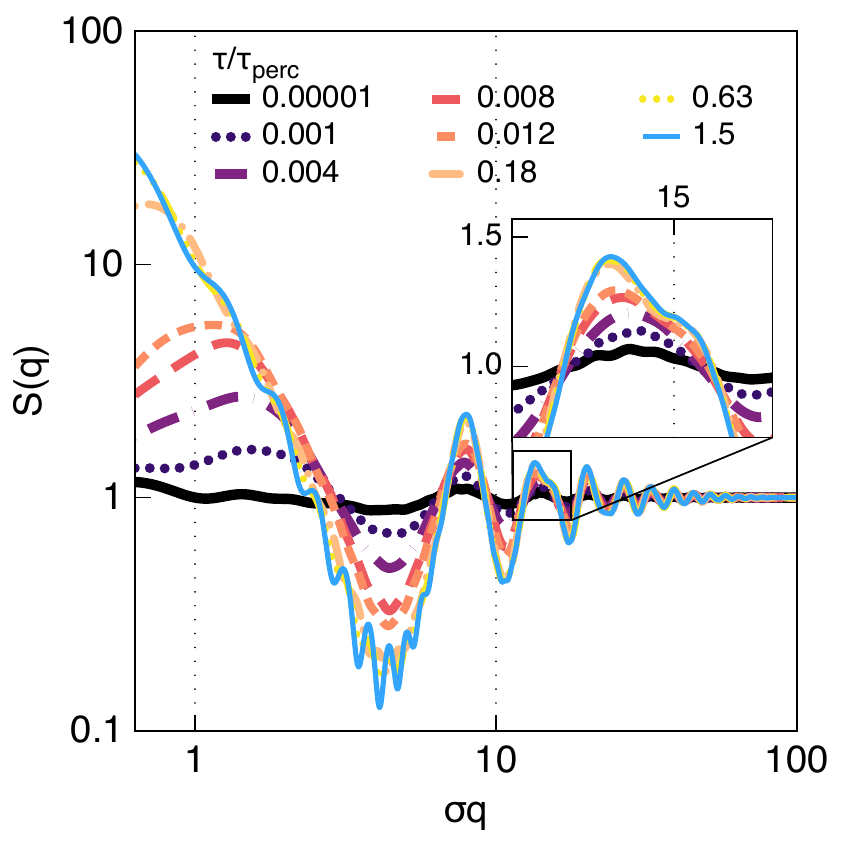}
	\caption{Double logarithmic plot of the time evolution of the structural changes as detected by the structure factor $S(q)$ (Eq.\ref{eq:sq}) for a system at volume fraction $\phi=0.05$ and interaction strength $\beta\varepsilon=10$. At early times the system is in a disordered gaseous state; when approaching the percolation time  $\tau_{\rm perc}$ and gelation, the magnitude of the low-q peak increases dramatically. Inset: local structure emerges in the shoulder of the peak at $q\sigma\approx 4\pi$. }
	\label{fig:grs}
\end{figure}

The process of formation of a percolating gel is accompanied by a complex structural organisation of the local neighborhood of the particles and, at a longer range, of the connectivity properties of the cluster. At the level of the two-point correlation functions, this is illustrated in Figure \ref{fig:grs}, where we show the time evolution of the structure factor $S(q)$ for a system at moderately low volume fraction and interaction strength. This displays a series of remarkable features: (i) the striking increase of the low-q peak illustrates the rapid formation of large, system-spanning structures when approaching the percolating regime; (ii) the shape of the peaks at $q\sigma\approx 4\pi$ indicates the formation of local order at times as early as $0.008$ percolation times; (iii) finally, the increase in amplitude of the secondary peaks and the emergence of longer-length scale oscillations suggest that the presence of medium-range correlations in the final percolated structures.

\subsection{Limits of gelation}
\label{subsec:limits}

\begin{figure}[b]
	\centering
	\includegraphics{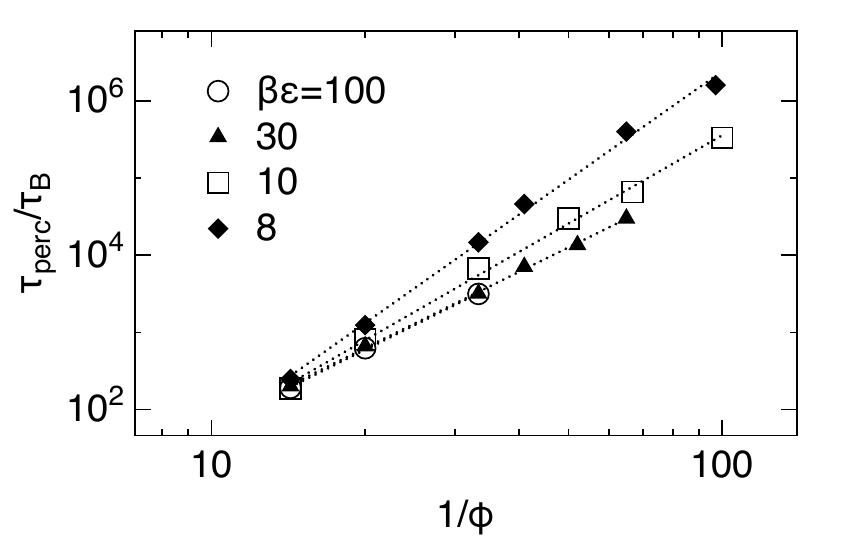}
	\caption{Elapsed time $\tau_{perc}$ before the detection of a percolating cluster as a function of the inverse volume fraction and the interaction strength in a double logarithmic plot. Dotted lines are power-law fits.}
	\label{fig:tauperc}
\end{figure}

It has been suggested in a previous study \cite{royall2015} (where the observation time was fixed in order to match the experimental conditions) that percolating gels in simulations without hydrodynamics become hard to access for volume fractions below $\phi^\ast\sim0.07$. Performing more extensive numerical simulations, we quantify this effect estimating the time needed in order to form a percolating cluster as a function of both the volume fraction and the interaction strength. As illustrated in Fig.~\ref{fig:tauperc}, the percolation time $\tp$ increases \paddyspeaks{by} several orders of magnitude when the volume fraction is reduced. Larger interaction strengths tend to reduce the time needed to percolate, and this effect is amplified at smaller volume fractions.

In particular, for moderate interaction strengths, we collect data down to low volume fractions $\phi\approx 0.01$. In the accessible dynamical range, the relation between percolation times and volume fraction appears to be governed by a power low $\tau_{\rm perc}\propto \phi^{-\alpha}$, with $\alpha(\BE)\in[3.3,4.7]$, suggesting that no characteristic or limit volume fraction can be detected. This also implies that any limit to the detection of a percolating cluster is mainly set by the observation time: if this is long enough, one can expect to observe a gel state even in the limit of extremely diluted suspensions. Similarly to inferences from experiments in microgravity\cite{manley2004}, {we rationalise our results observing} that gelation is fundamentally limited only by the magnitude of thermal fluctuations, capable of dissolving the percolating clusters.

 \begin{figure}[t]
	\centering
	\includegraphics{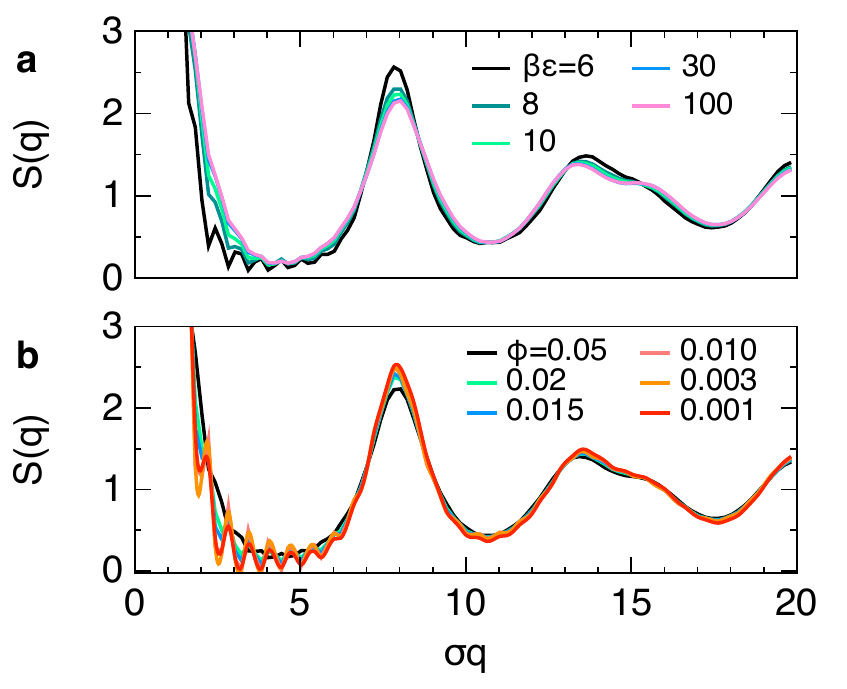}
	\caption{Structure factor $S(q)$ for (a) fixed volume fraction $\phi=0.05$ and several values of the interaction strength and (b) fixed interaction strength $\BE=10$ and varying volume fraction. The small oscillations at $\sigma q<5$ are an artefact of Eq.\ref{eq:sq}.}
	\label{fig:sqs}
\end{figure}

\subsection{Structural properties at low volume fractions}
\label{subsec:low}

Once a percolating cluster has formed, we analyse the structural properties of the system in order to identify the nature of the different gels obtained at different thermodynamic conditions. 
First we focus on the structural changes as detected by the structure factor.

In Fig.~\ref{fig:sqs}(a) we illustrate the effect of changing the interaction strength on a system that is moderately dense, $\phi=0.05$: the characteristic wavelengths are unchanged by the changing interaction strength; however, one observes that for ($\BE>6$) the peak at $\sigma q\approx 4\pi$ presents a shoulder, indicative of the increase of local order. At the same time, the peak at $\sigma q\approx 2\pi$ decreases in height, accompanied by the increase in amplitude corresponding to longer-range wavelengths.
 
In Fig.~\ref{fig:sqs}(b) a similar scenario is obtained when lowering volume fractions at a fixed value of the interaction strength $\BE=10$. However, the increase of local order as detected at wavelengths close to $4\pi$ is accompanied by a further increase in the amplitude of the peak at $2\pi$ and an overall depletion of the long wavelength modes. These results indicate that the change of structure hinted by the shoulder $\sigma q\approx 4\pi$ takes a different form if one decreases the volume fraction or increases the interaction strength.
%
%

\begin{figure}[t]
	\centering
	\includegraphics{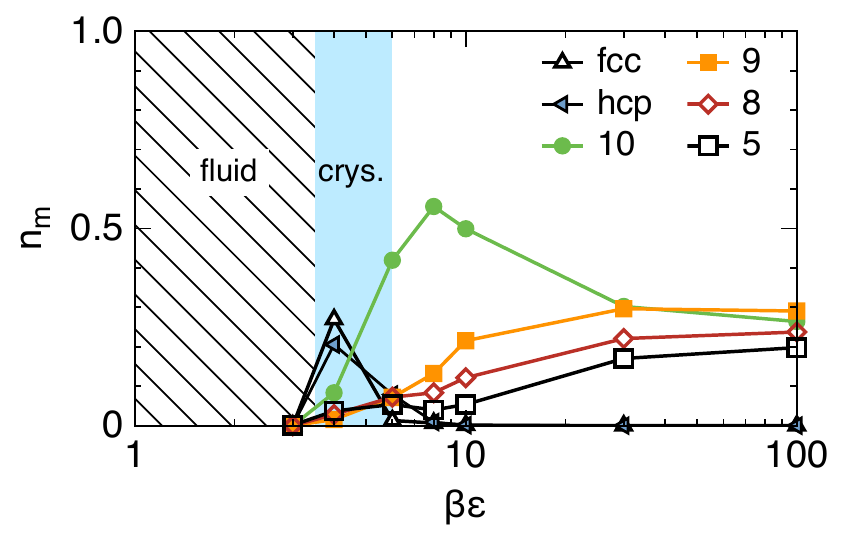}
	\caption{Concentration $n=N_m/N$ of particles identified in local structures of $m$ particles at $\phi=0.03$. A disordered fluid phase at low $\BE<3.5$ is followed by a phase where crystalline order prevails ($3.5<\BE<6$), and finally a percolating phase dominated by clusters of size $m=10$, formed by five-membered rings. Error-bars are within the size of the symbols.}
	\label{fig:tccphi05}
\end{figure}

\begin{figure}[t]
	\centering
	\includegraphics{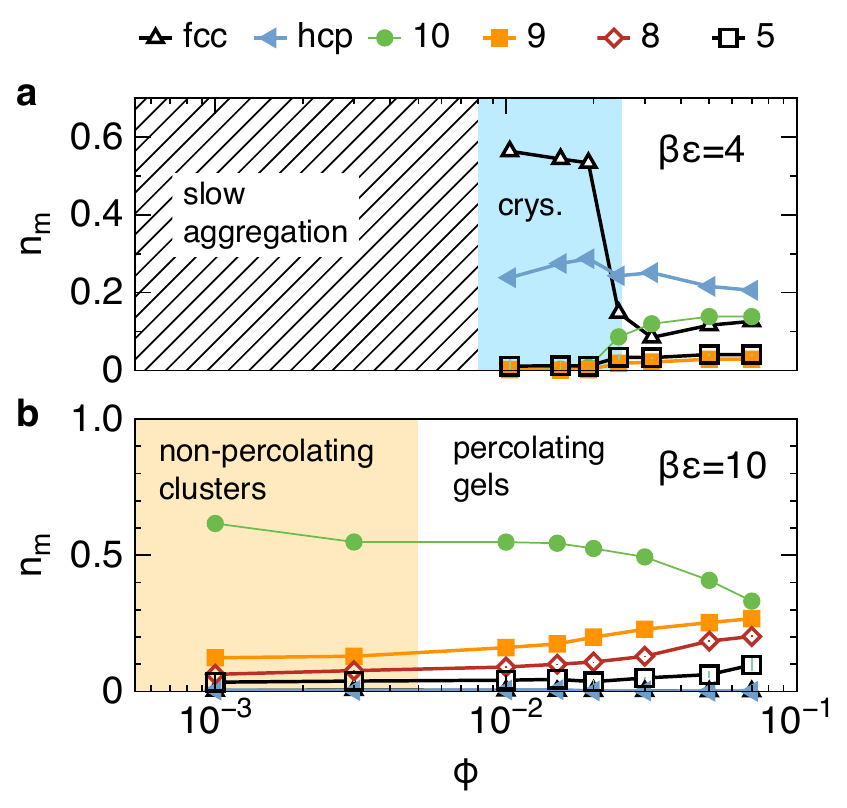}
	\caption{Concentration $n_m=N_m/N$ of particles identified in structures of $m$ particles at $\BE=4$ (a) and $\BE=10$ (b). In (a) crystalline clusters, dominated by fcc local order, are formed for $\phi<0.025$. In (b) percolating gels ($\phi>0.005$), and non-percolating aggregated clusters ($\phi<0.005$) are formed. Only relevant structures are shown. Errorbars are within the symbol sizes.}
	\label{fig:tccbe10}
\end{figure}

In order to quantify the nature of the different behaviour, we compute higher order correlations. More detailed knowledge on the kind of local order appearing in the formation of the percolating gels can be obtained from the Topological Cluster Classification analysis of the particle coordinates (see \ref{sec:metric}). This method identifies, within the network of neighbours, domains composed by arrangements of particles compatible with the local minima of $m$ particles interacting via the Morse potential. The results of this analysis are shown in Figures \ref{fig:tccphi05} and \ref{fig:tccbe10} for constant volume fraction and constant interaction strength conditions respectively {\cite{malins2013tcc}}.

Among the several structures identified by the TCC, some specific arrangements have a prominent role, depending on the interaction strength and the volume fraction. At fixed volume fraction $\phi=0.03$ (Fig.~\ref{fig:tccphi05}), we observe that starting from very high interaction strength $\BE$ the particles (kinetically arrested in their initially random relative positions) tend to form  structures compatible with $m=8,9$ and 10 {which all are five-fold symmetric polyhedra}. The {$m=10$ defective icosahedra} 
in particular, become more and more represented as the interaction strength is reduced while lower order structures (5 to 9 particles) are less present, indicating that aggregation is more accessible and larger low energy structures can be formed. This is even more evident when the interaction strength is decreased below $\BE=8$ where local order is enhanced in the form of local crystalline structures such as face centered cubic or hexagonal close packed structures,{consistent with previous work in the present \cite{royall2012} and related \cite{klotsa2011} systems.} We notice that the crystalline clusters form separately and at such low densities they consist in isolated clusters that do not percolate through the system. For $\BE\lesssim 3$ the attractions are so weak that the thermal fluctuations are sufficient to stabilise a fluid phase, and make gelation impossible. 

\begin{figure}[t]
	\centering
	\includegraphics{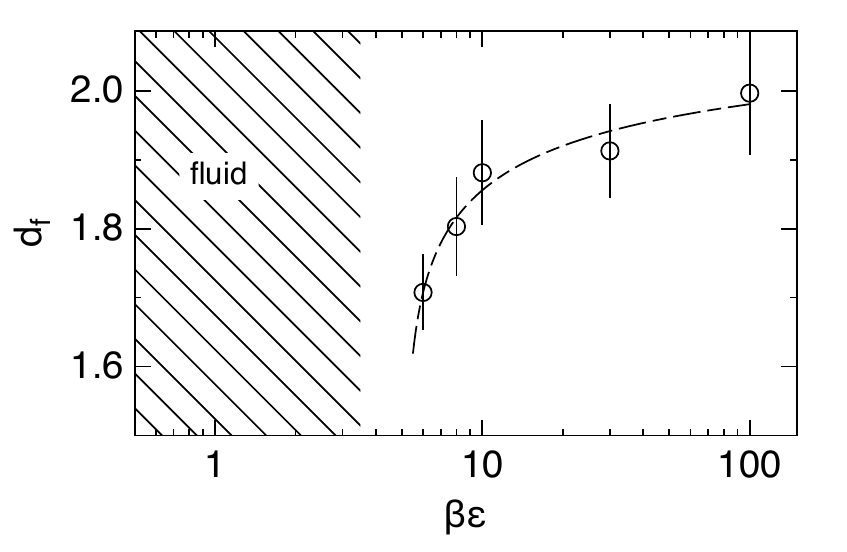}
	\caption{Fractal dimension as a function of the interaction strength  $\beta\epsilon$ at volume fraction $\phi=0.05$. The shaded area indicates the region for which percolation is not observed. The dashed lines serves as a guide to the eye.}
	\label{fig:fracdim}
\end{figure}

\begin{figure}[t]
	\centering
	\includegraphics{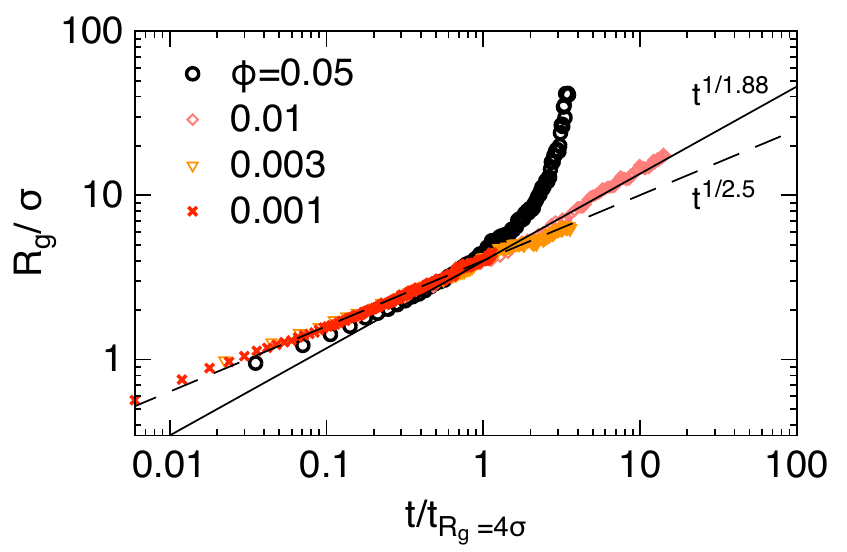}
	\caption{Evolution of the average radius of gyration of the aggregates for different volume fractions at $\BE=10$. For $\phi=0.05$ a percolating cluster is actually observed within the simulation time, hence the sheer increase in the average radius of gyration. The continuous line indicates the slope of $t^{1/d_f}$ where $d_f$ is estimated for $\phi=0.05$, while the dashed line corresponds to $d_f=2.5$, associated to diffusion-dominated cluster accretion. Time is rescaled by $t_{R_g=4\sigma}$, the time at which the average radius of gyration is equal to $4\sigma$.}
	\label{fig:radii}
\end{figure}

\begin{figure*}[t]
	\centering
	\includegraphics[width=\textwidth]{{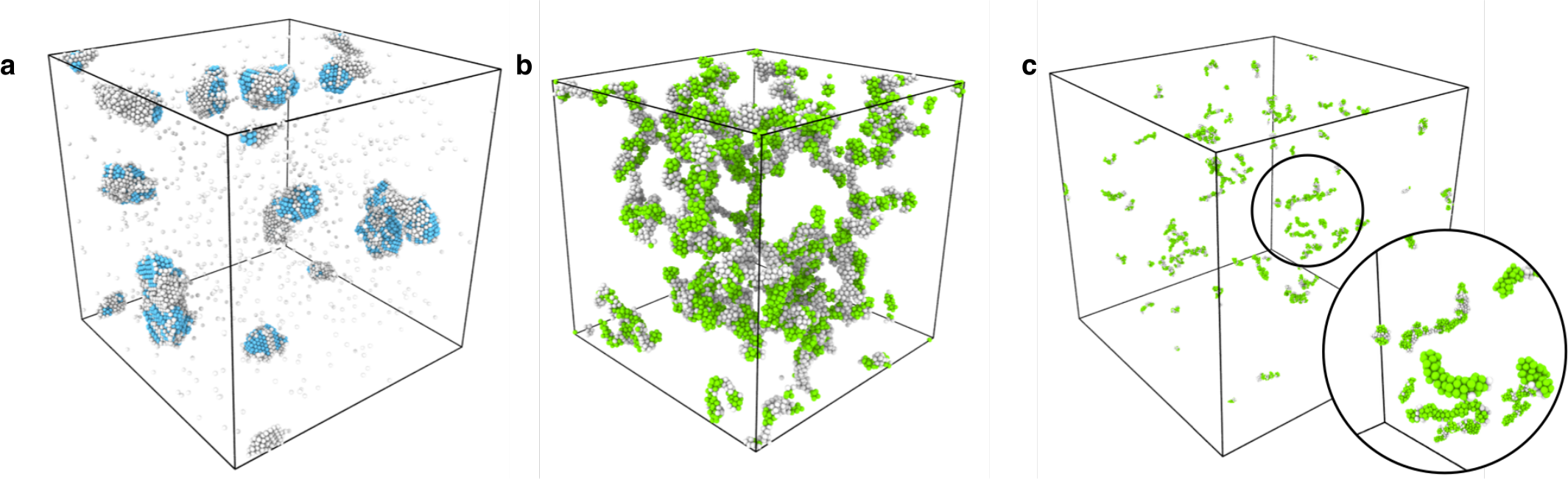}}
	\caption{ Final configurations at different volume fractions and interaction strengths: (a) a crystalline cluster phase at $\BE=4$ and $\phi=0.0154$ {with ``crystal droplets''}; (b) a gel at $\BE=10$ and $\phi=0.03$; (c) a cluster phase at $\BE=10$ and $\phi=0.001$. Phases (b) and (c) are dominated by the $m=10$ defective icosahedra local order (green) while (a) shows local fcc crystalline order (azure) coexisting with a gaseous phase.}
	\label{fig:snapstcc}
\end{figure*}

\begin{figure}[tb]
	\centering
	\includegraphics[width=0.6\columnwidth]{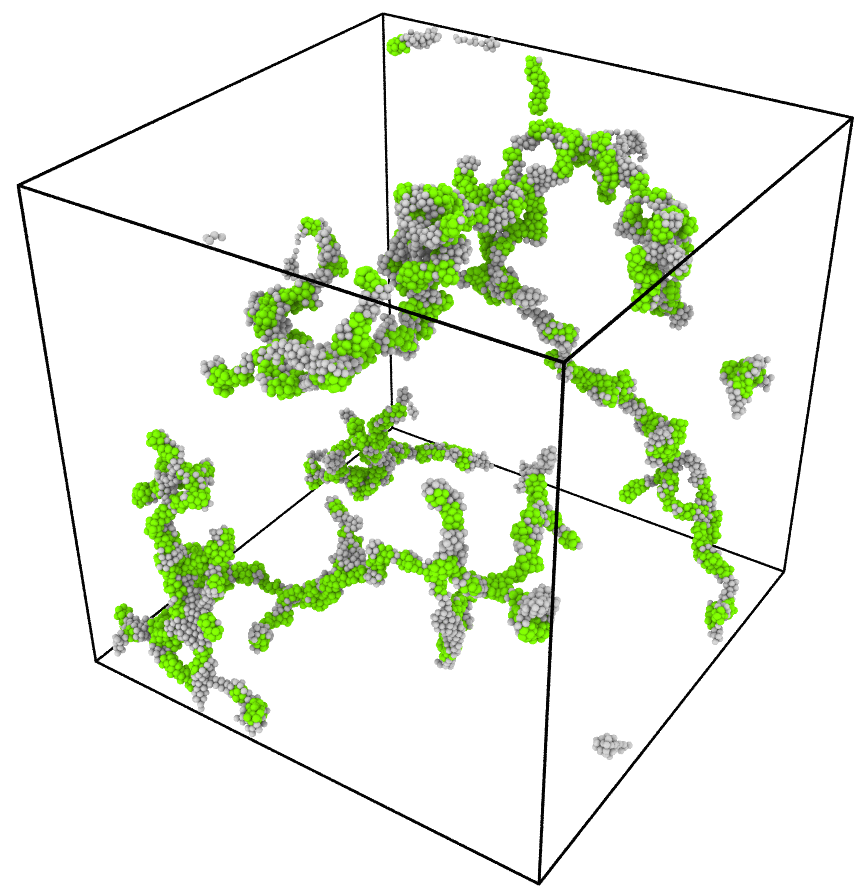}
	\caption{ The particles in the local structure corresponding to the $m=10$ Morse energy minimum (green) form the backbone of the percolating gel filaments at very low volume fractions (here $\phi=0.01$) at $\BE=10$. }
	\label{fig:10Bphi01}
\end{figure}

We then analyse two exemplary cases at fixed interaction strengths and variable volume fraction: $\BE=4$ (compatible with local crystalline order) and $\BE=10$ (where crystals are rare), see Fig.~\ref{fig:tccbe10}. For $\BE=4$ we find that the density fluctuations are sufficient to trigger the nucleation of local crystalline clusters. When the volume fraction is above $\phi=0.025$, there is a variety of local arrangements in hcp, fcc or five-fold symmetric order ($m=10$), but below $\phi=0.025$ the fcc local order dominates and the system is mainly formed by isolated crystalline clusters immersed in a gaseous phase (see Fig.~\ref{fig:snapstcc}(a)), {like ``droplets of crystals''}. {For values of $4<\BE<10$ we observe an intermediate regime where crystalline and noncrystalline structures contribute to the formation of a percolating network.}

For a larger interaction strength ($\BE=10$), we vary the volume fraction and track the structural changes in the formed gels,  Fig. \ref{fig:tccbe10}(b). The kind of structures formed ranges from percolating gels to a cluster phase, as depicted in Fig. \ref{fig:snapstcc}(b-c). In this case, we see that the fraction of particles participating to low energy structures such as the $m=10$ defective icosahedra (indicative of five fold symmetry) increase systematically, at the expense of smaller clusters as the volume fraction $\phi$ is reduced. This suggests that when the gel is made only by narrow filaments due to the paucity of the particles available, these are mainly arranged into low energy structures, which form the backbone of the percolating network, see Fig.~\ref{fig:10Bphi01}. The deep energy minima represented by the $m=10$ defective icosahedra are then responsible of the mechanical stiffness of the gel and its resistance to thermal fluctuations. It is important to notice that it is precisely this kind of structures that are underrepresented in experimental conditions\cite{royall2008,royall2014arxiv,royall2015}, due to the effects of hydrodynamic interactions, excluded from the present study. 
\\

All the previous measurements are local to some degree. In order to quantify how the percolation network changes globally as we change the interaction strength we measure a global property of the network, as its fractal dimension. We perform a measure of structural order through the estimation of the average fractal dimension $d_f$ of the percolating clusters via box counting (as described in section \ref{sec:metric}). In Figure \ref{fig:fracdim} we show how, for a moderately dilute system at $\phi=0.05$, the change in the interaction strength leads from space-filling percolating networks for high $\BE$  ($d_f\sim2$) to more chain-like structures at lower $\BE$, as hinted by the decay of $d_f$ towards $d_f\sim1.7$ when approaching the gel-fluid boundary at $\BE\approx 3.5$. These results are consistent with the fact that diffusion-controlled cluster accretion \cite{meakin1983}(where single diffusing particles coalesce on a seed) is compatible with fractal dimension $d_f\sim2.5$ in three dimensions while cluster formation by diffusion-limited aggregation \cite{weitz1984}  (where clusters of comparable size aggregate) is compatible with $d_f\sim1.75$. The simple, single particle diffusive motion is therefore the dominant mechanism at very high interaction strengths while coalescence of {equal size} chain-like clusters prevails at smaller $\BE$.


\subsection{Low volume fractions}
\label{subsec:xlow}

We now consider even lower volume fractions and discuss the structural properties of the aggregates that we obtain in the light of the percolating gels illustrated in the previous section.

When decreasing the densities to extremely low values ($\phi=0.003,0.00$1), the time needed to form a percolating clusters exceeds the available computation time (the longest calculations lasted 2688 CPU hours). The evolution of the system proceeds through a slow ggregation that nevertheless permits the formation of local, low energy aggregates. 
In Figure \ref{fig:tccbe10} we demonstrate two distinct behaviours, as a function of the interaction strength: for moderate interaction strengths $\BE=4$, Figure \ref{fig:tccbe10}(a), we observe that the local crystalline order prevails down to very low packing fractions. In particular, for $\phi<0.025$, a large fraction of particles resides in an fcc-like ordered cluster, coexisting with a very dilute gas of isolated particles. {For volume fractions even smaller ($\phi<0.01$) the time necessary for aggregation is even slower and it exceeds the accessible computation time}.

Conversely, when the interaction strength is strong enough, the crystalline order is largely suppressed. As demonstrated in Figure \ref{fig:tccbe10}(b), the structural signature of these disconnected clusters is consistent with the features of higher density percolating gels, indicating that at the local level it is hard to distinguish percolating from non-percolating systems. In particular, we notice that the $m=10$ geometry still dominates the statistics, showing that the aggregation stems from the formation of low energy clusters of five-fold-membered rings.

{At such low volume fractions, forming a percolating network becomes challenging: nonetheless we can compare the typical structure of the aggregates tracking the time evolution of the radius of gyration $R_g$ as a function of time (Fig.~\ref{fig:radii}). Diffusion-limited cluster aggregation predicts that $R_g\sim t^{1/d_f}$ where $d_f$ is the fractal dimension of the cluster \cite{weitz1985,carpinetti1992}. For volume fractions ranging from $0.001$ to $0.05$ at a fixed interaction strength $\BE=10$ we observe that while at very early times $d_f\approx 2.5$, at later times the growth of the radius of gyration is compatible with $d_f\approx1.88$, the fractal dimension of the percolating gel at $\phi=0.05$ determined by box counting. The data collapse suggests that the growth mechanism is the same for the same interaction strength, and that the low volume fractions have the main consequence of dramatically slowing down the aggregation process.}

\section{Conclusions}

By the means of numerical simulations, we have explored the formation of model nonequilibrium colloidal gels in the limit of very low {volume fractions} and high interaction strengths in a solution modelled by over-damped Langevin dynamics. The considered state points, Fig.~\ref{diagram}, allowed us to demonstrate that, under the idealised conditions of our numerical simulation, percolating gels can be found down to very low densities, provided that the observation time is sufficiently long. In fact, we have shown that the time necessary in order to observe a percolating cluster increases rapidly with the inverse volume fraction, suggesting that density alone cannot be a limit to gel formation.

We have also explored extremely low volume fraction gels, where percolation occurs on 
longer time scales than we can access. Comparing the structural features of these extremely low density aggregates and the higher density percolating networks we observe very similar patterns, suggesting that the mechanisms at play at the local level are the same:  rigidity emerges from the condensation of the colloids into locally favoured structures \cite{royall2008}. In particular, we observe that five-fold symmetric order (represented by the $m=10$ {defective icosahedra}) plays a very important role in forming the backbone of the chain-like network realised at very low densities.

On the basis of this analysis, we {extended} previous phase diagrams \cite{royall2015}, identifying a new region where Brownian dynamics simulations can realise percolating gels if the observation time is longer than the percolation time (Fig.~\ref{diagram}): the obtained gels are however strikingly different from the structural point of view with respect to those found in experiments \cite{royall2008,royall2014arxiv,royall2015}, with more compact aggregates and prevalence of five-fold symmetric order. This  emphasises the importance of (here absent) long range, hydrodynamic interactions in order to form less compact structures \cite{furukawa2010,royall2015}. 

{The overall physical scenario suggests that gelation occurs when two necessary but not individually sufficient conditions are satisfied (see Fig.~\ref{diagram}): on one hand, the interaction strength needs to be large enough to drive phase separation to a liquid and gas; on the other hand, the observation time needs to be sufficiently long to allow for percolation to occur. In fact, even when demixing occurs at low volume fractions for too weak interaction strengths, the resulting liquid droplets -- which may then crystallise -- appear too compact to percolate if $\BE\lesssim 8$. In the case of very low volume fractions, very thin filaments are expected to be formed, which appear as isolated clusters for observation times shorter than the percolation time. However, the local structure of such filaments would be hardly distinguishable from the local structure of the percolating network.}

Building on the present results for the purely Brownian case, it will be possible to understand how the hydrodynamic {forces} affect the dynamics (in terms of the time required to form a percolating cluster), the local order and the connectivity properties of the network in the limit of very low densities once the hydrodynamic effects will be added ({in the form of, for example,} multi-particle collision dynamics \cite{gompper2009multi}). This will {contribute} to a microscopic explanation of the emergent rigidity in colloidal gels.

\bibliography{connection_sub}
\end{document}